\documentclass{aastex}
\usepackage{emulateapj5}

\def\lesssim{\mathrel{\hbox{\rlap{\hbox{\lower4pt\hbox{$\sim$}}}\hbox{$<$}}}}
\def\gtrsim{\mathrel{\hbox{\rlap{\hbox{\lower4pt\hbox{$\sim$}}}\hbox{$>$}}}}

\shorttitle{Environmental Effects on Cluster Galaxies}
\shortauthors{Okamoto \& Nagashima}

\begin{document}


\title{Environmental Effects on Evolution of Cluster Galaxies 
in a $\Lambda$CDM Universe}


\author{Takashi Okamoto}
\affil{Department of Physics, University of Durham, South Road, Durham DH1 3LE England
and\\
Yukawa Institute for Theoretical Physics, Kyoto Univ., Sakyo-ku, Kyoto 606-8502 Japan\\}
\email{Takashi.Okamoto@durham.ac.uk}

\and

\author{Masahiro Nagashima}
\affil{National Astronomical Observatory, Mitaka, Tokyo 181-8588 Japan}
\email{masa@th.nao.ac.jp}


\begin{abstract}
We investigate environmental effects on evolution of bright cluster
galaxies ($L > L_*$) in a $\Lambda$ dominated cold dark matter universe
using a combination of dissipationless $N$-body simulations and a
semi-analytic galaxy formation model.  The $N$-body
simulations enable us to calculate orbits of galaxies in simulated
clusters.  Therefore we can incorporate stripping of cold gas from
galactic disks by ram pressure (RP) from intracluster medium into our
model.  In this paper we study how ram pressure stripping (RPS) and
small starburst induced by a minor merger affect colors, star formation
rates (SFRs), and morphologies of cluster galaxies. These processes are new
ingredients in our model and have not been studied sufficiently.  We
find that the RPS is not important for colors and SFRs of galaxies in the
cluster core if star formation time-scale is properly chosen, because
the star formation is sufficiently suppressed by consumption of the cold 
gas in the disks.  
Then observed color and SFR gradients can be reproduced without the RPS.  
The small starburst triggered by a minor merger hardly affects the SFRs
 and colors of the galaxies as well.  
We also examine whether these two processes can resolve the known 
problem that the hierarchical clustering models based on the major 
merger-driven bulge formation scenario predict too few galaxies
of intermediate bulge-to-total luminosity ratio ($B/T$) in clusters.
When the minor burst is taken into account, the intermediate $B/T$
population is increased and the observed morphology gradients in
clusters are successfully reproduced. 
Without the minor burst, the RPS cannot increase the intermediate $B/T$ 
population. On the other hand, When the minor burst is considered, 
the RPS also plays an important role in formation of the intermediate 
$B/T$ galaxies. 
We present redshift evolution of morphological fractions predicted by
our models.  The predicted number ratios of the intermediate $B/T$
galaxies to the bulge-dominated galaxies show nearly flat or slightly
increasing trends with increasing redshift.  We conclude that these 
trends are inevitable when bulges are formed through mergers.  We discuss
whether our results conflict with observationally suggested $N_{\rm
S0}/N_{\rm E}$ evolution in clusters, which is a decreasing function of
redshift.

\end{abstract}

\keywords{galaxies: formation --- galaxies: evolution --- %
galaxies: clusters: general --- galaxies: %
halos --- galaxies: interactions}

\section{INTRODUCTION}

Rich clusters of galaxies are the most suitable laboratories to study how
galaxy environments affect galaxy properties.  The pioneering work by
Butcher \& Oemler (1978, 1984) has demonstrated that distant clusters
contain much higher fraction of blue galaxies than nearby clusters, 
though it has been questioned by \citet{and99}.  
It has been also found that galaxy morphology is a function of its
environment \citep{dre80, whi93}.  The most recent advances have been
made with the {\it Hubble Space Telescope}, which allows the morphology
of distant galaxies to be directly compared with the properties of their
nearby counterparts.  \citet{dre97} and \citet{cou98} have suggested
that the predominant evolutionary effects are that the distant clusters
have a substantial deficit of S0 systems compared with nearby clusters,
and at lower luminosities they contain substantial Sc-Sd spirals,
compared with the large population of dwarf spheroidals in present-day
clusters, while it has also been questioned by \citet{and98}. 
On the other hand, some recent studies \citep[e.g.][]{abr96,
vanD98, bal99} have focused on observed trends in star formation rate 
(SFR) and morphology as a function of position within a cluster.  
These studies have shown that there is smooth transition from a blue,
disk-dominated population of galaxies in the outskirts of clusters to a
red, bulge-dominated population in the cluster cores.

Many authors have suggested that the predominance of red, early-type
galaxies in local clusters is the result of mechanisms that suppress
star formation in high density environments.  This suppression makes
galaxies within cluster cores redder and leads to a transformation of
galaxy morphology.  Comparison between the galaxy populations of local
and distant clusters or those in cluster cores and outskirts provides a
strong evidence for this scenario.  This leads to a naive conclusion
that the primary effect of the cluster environment is to transform
luminous spiral galaxies into S0 galaxies through the suppression of
their star formation.

Several mechanisms that may suppress the star formation and transform
one morphological type into another have been proposed.

Interactions between galaxies are one possible process to promote
morphological transformation. $N$-body simulations confirmed that major
mergers between disk galaxies produce galaxies resembling ellipticals as
merger remnants \citep[e.g.][]{bar96} and that accretion of small
satellites onto their host spiral lead the morphology of the host spiral
to S0 type \citep{wal96}.  The galaxy mergers trigger the {\it
starburst}, therefore the cold gas contained in these galaxies are exhausted
in a very short time. These processes then cause the truncation of star
formation.  Although the bulge formation scenario by major mergers gives
an excellent explanation both for colors \citep{kc,ng01} and
distribution of cluster ellipticals \citep{on01, dia01}, the observed S0
population cannot be reproduced only by major mergers \citep{on01,
dia01}.  In addition, the galaxy mergers are predominant effect before
massive cluster formation, because, in clusters, the relative velocity
of galaxies is too high for such mergers to be frequent
\citep{ghi98,oka99}. Therefore the galaxy merger is not promising as a
key mechanism behind the Butcher-Oemler effect.  \citet{moo96} examined
the effects of rapid gravitational encounters between galaxies. This
mechanism has been called {\it galaxy harassment} and is highly
effective at transforming fainter Sc-Sd galaxies to dSphs.  While the
galaxy harassment can account for the observed evolution of faint
galaxies in clusters, the concentrated potentials of luminous Sa-Sb
galaxies help to maintain their stability, then the bulge-to-disk
luminosity ratios ($B/T$s) of these
galaxies are hardly changed only by the galaxy harassment \citep{moo99}.
Nevertheless, if the truncation of star formation in these galaxies
takes place, these galaxies can become the objects that are similar to
S0's, because the disks of Sa-Sb galaxies are substantially thickened
and the spiral features are vanished by the harassment.  It should be
noted that correlation between $B/T$ and Hubble type has significant
scatter \citep{bau96} and there is no guarantee that this relation is
applicable to cluster galaxies.  Accordingly we adopt the classification
based on $B/T$ to observed galaxies as well as to our model galaxies.

The second mechanism is the truncation of star formation through the
removal of the diffuse hot gas reservoir that is confined in galactic
dark halos and surrounds galaxies \citep{lar80}. In clusters, any hot
diffuse material originally trapped in the potentials of the galactic
halos becomes part of the overall intracluster medium
(ICM).  A galaxy whose hot gas reservoir is removed slowly exhausts its
cold gas in a few gigayears \citep[e.g.][]{gal89}, because there is no
supply of the fresh gas from the surrounding hot gas.  This process is
an important ingredient in the success of semi-analytic (SA) models,
which have been shown to reproduce observed global properties of cluster
galaxies such as the morphological composition and the blue galaxy 
fraction \citep[e.g.][]{bau96, kau96}.  
Although such slow truncation after the infall into larger virialized 
halos explains the observed SFR gradients in clusters \citep{bal00} and 
the color evolution of cluster galaxies \citep{kb01}, 
the lack of the population of intermediate $B/T$ cannot be solved even
by including this mechanism into the galaxy formation model \citep{dia01}.

The other mechanism is ram pressure stripping (RPS) of the cold gas from 
galactic disks \citep{gun72}.  
As a galaxy orbits through a cluster, it feels ram pressure (RP) from 
the ICM. When the RP is greater than the binding force, 
the cold gas will be stripped in $\sim 10^7$ yr \citep{aba99}.  
\citet{fn99} suggested that the RPS increases $B/T$ of 
the Milky Way-like galaxies owing to the rapid suppression of
the star formation in the disk component, and consequently Sb galaxies
change their morphology into S0 galaxies by assuming static cluster
potentials and radially infalling galaxies.

In this paper we investigate how these mechanisms affect colors and
morphologies of cluster galaxies with a special interest in the
effects of the RPS and the minor mergers that have not been well 
studied in cosmological context.  
Here we do not attempt to model the galaxy harassment, because we only 
treat bright galaxies ($L > L_*$) and the galaxy harassment probably do 
not affect strongly for such bright galaxies as far as we classify the 
galaxies by $B/T$ as we mentioned previously. 

For this purpose, the fully numerical simulations are still too
expensive and have a lot of difficulty in handling {\it subgrid}
physics, e.g. gas cooling, star formation, supernova feedback, chemical
evolution and so on, though recent work is beginning to achieve some
notable successes for the distribution of galaxy populations
\citep[e.g.][]{sm95, wei97, bla99, khw99, pea99, co00}.

Alternative method to study the evolution of galaxy population is the SA
modeling of galaxy formation \citep[e.g.][]{kau93, col94, som99, col00,
nag01, nag02}, which follows the collapse and merging of dark halos by
using a probabilistic method on the mass distribution based upon an
extension of the Press-Schechter formalism \citep{pre72, bon91, bow91,
lac93} and in which the subgrid physics is incorporated with the merging
histories of dark matter halos assuming simple scaling laws.

As a natural expansion of this approach, hybrid methods of cosmological
$N$-body simulations and the phenomenological model of galaxy formation
used in the SA techniques have been developed \citep{rou97, kau99,
ben00}.  
In order to obtain the spatial and velocity distribution of 
galaxies and to treat the mergers between galaxies in cluster
environment, several authors have adapted the refinements of hybrid
methods, which traces the merging histories of substructures within
virialized halos \citep{on01, spr01}.  Since this method enables us to
calculate orbits of galaxies in clusters, we can incorporate the RPS in our
model as well as merging of substructure halos.

We compare our simulated cluster at $z = 0.2$ to an observed sample that
is constructed by superposing 7 low redshift CNOC1 \citep{yee96}
clusters ($0.18 < z < 0.3$).  To compare our results with the
observations, we rescale our simulated cluster and the CNOC1 clusters by
$R_{200}$ which is the radius of the sphere centered on the cluster
center and whose density is 200 times the critical density of the
universe at a given redshift.

In \S 2, we describe an outline of the galaxy formation model used here
and we show the characteristics of our model in \S 3.  Results on the
luminosity functions, star formation gradients, color gradients, and
morphological gradients are presented in \S 4.  These results are
discussed in \S 5.

\section{MODEL}

We examine the evolution of  cluster galaxies in a $\Lambda$CDM universe
($\Omega_0  = 0.3$,  $\lambda_0 \equiv \Lambda_0/(3 {H_0}^2) = 0.7$,  
$h \equiv  H_0/100$ km s$^{-1}$ Mpc$^{-1} =  0.7$, $\sigma_8 = 1.0$).  
The  baryon density, $\Omega_{\rm b}$, is set to $0.015 h^{-2}$
(Suzuki, Yoshii, \& Beers 2000). 
Note that a recent measurement of the anisotropy of the cosmic microwave
background suggests a slight higher value, 
$\Omega_{\rm b} \simeq 0.02 h^{-2}$ \citep{net02}.  
The changing of $\Omega_{\rm b}$ affects only the normalization of 
parameters in our SA model and we confirmed that it does not affect our
conclusion. Thus, we here prefer using $\Omega_{\rm b} = 0.015 h^{-2}$ 
that was adopted in our previous paper \citep{on01}. 

The outline of the procedures of galaxy formation is as follows.  At
first, the merging paths of dark halos are realized by a cosmological
$N$-body simulation. At this stage, it is determined for each halo
whether it is a central halo of its virialized host halo or it is a
substructure of its host.  Next, in each merging path, evolution of the
baryonic components, namely, gas cooling, star formation, supernova
feedback, and chemical evolution, are calculated based on the SA 
galaxy formation model.  
The detail of our SA model will be given in \S \ref{SAmodel}. 

We refer to a system consisting of the stars and
cold gas as a {\it galaxy}.  
When two or more galactic halos merge together, 
we estimate merging time-scale of the galaxies in the new common halo 
based on dynamical friction time-scale.  
As merging of galaxies occurs, we change the morphology of
the merger remnant according to the type of the merger.  Finally, we
calculate the luminosity and color of each galaxy based on the stellar
populations that compose the galaxy.

\subsection{Simulations} \label{simulation}

We first describe two simulations used in this paper and their specific
purpose. These simulations are the same as those by  Okamoto \&
Habe (1999, 2000) except for the background cosmology.

For a main simulation, we adopt the constrained random field method to
generate the initial density field in which a rich cluster
is formed at the center of a simulation sphere of radius 22.5 $h^{-1}$
Mpc \citep{hr91}.  The constraint that we
impose is the 3 $\sigma$ peak with the 8 Mpc Gaussian smoothed density
field at the center of the simulation sphere. We refer this simulation
as the {\it cluster simulation}.

Another simulation represents an average piece of the universe
corresponding to the {\it field} environment within a sphere of radius
20 $h^{-1}$ Mpc without any constraints.  We use this simulation to
normalize the parameters in our galaxy formation models and to compare
the galaxy population to that in the cluster simulation.
Although the size of this simulation seems to be somewhat small, 
\citet{oka99} obtained a reasonable halo mass function from a simulation 
sphere of radius $3.5 h^{-1}$ Mpc. 
We have confirmed that this simulation produces a reasonable mass function 
and the mass of the most massive halo at $z = 0$ is sufficiently small 
($1.9 \times 10^{13} M_{\odot}$) as a sample of low-density environment.

To get sufficient resolution with relatively small number of particles
we use a multimass initial condition for each simulation.  In the field
simulation, we put the high resolution particles in the sphere of radius
10 $h^{-1}$ Mpc and in the remaining region we put the low resolution
particles as boundary particles.  On the other hand, we adopt the
resimulation technique for the cluster simulation \citep{oka99}. That
is, first, only long-wavelength components are used for the realization
of the initial perturbations in the simulation sphere using $\sim 10^5$
particles, and then we perform a simulation with these low resolution
particles. After this procedure, we tag the particles that are inside
the sphere of radius 3 $h^{-1}$ Mpc centered on the
cluster center at $z = 0$.  Next we go back to the initial redshift, and
then we divide the tagged particles according to the density
perturbations that is produced by additional shorter wavelength
components.  As a result, the number of the high resolution particles
becomes $\sim 10^6$.  We adopt the same mass and softening length
of a high resolution particle in each simulation. Our analyses are
operated only for these high resolution particles.

The overall parameters and mass of the most massive virialized object
in each simulation at $z = 0$ are listed in Table~\ref{tbl-1}.

\subsection{Halo Identification and Construction of Merger Trees} \label{mergertree}

The method to create merging history trees of dark halos is basically
the same as \citet{oka00}, while we make some improvement.

To construct the merger trees we identify virialized halos and their
substructures at 34 redshifts during $z = 20$ and $z = 0$.  
This procedure is as follows.

At each time-step, we calculate density field by smoothing each particle with 
neighboring 64 particles using a similar method to the smoothed particle 
hydrodynamics (SPH).  At a given redshift, particles with local densities greater 
than one-third of the virial density at that redshift are tagged.

We then operate the friends-of-friends (FOF) grouping algorithm 
\citep{dav85} only on the tagged particles, otherwise the FOF is likely
to find strangely-shaped halos like strings in low density regions 
\citep[][]{ber91}. 
The linking length is defined as $b \bar{l}$, where
$\bar{l}$ is the mean particle separation, and the parameter $b$ is a
function of redshift.  Its value is given by
\begin{equation}
b = \left(\frac{1}{2} \frac{\rho_{\rm vir}}{\bar{\rho}}\right)^{-\frac{1}{3}}, 
\label{link}
\end{equation}
where $\rho_{\rm vir}$ is the virial density calculated by the spherical
collapse model at a given redshift and $\bar{\rho}$ is the background
density of the universe. Note that Eq. (\ref{link}) implies density of 
a halo edge is half of the virial density that well exceeds the criterion 
used to tag the particles. 
This equation gives $b \simeq 0.2$ when we assume 200 as the virial 
overdensity, which is often used in the case of the critical density universe.  
When the FOF links more than 10 tagged particles, we identify this
linked particle group as a virialized object. 
We call the halos found by this procedure {\it FOF-halos}.

\begin{figure}
\epsscale{0.9}
\plotone{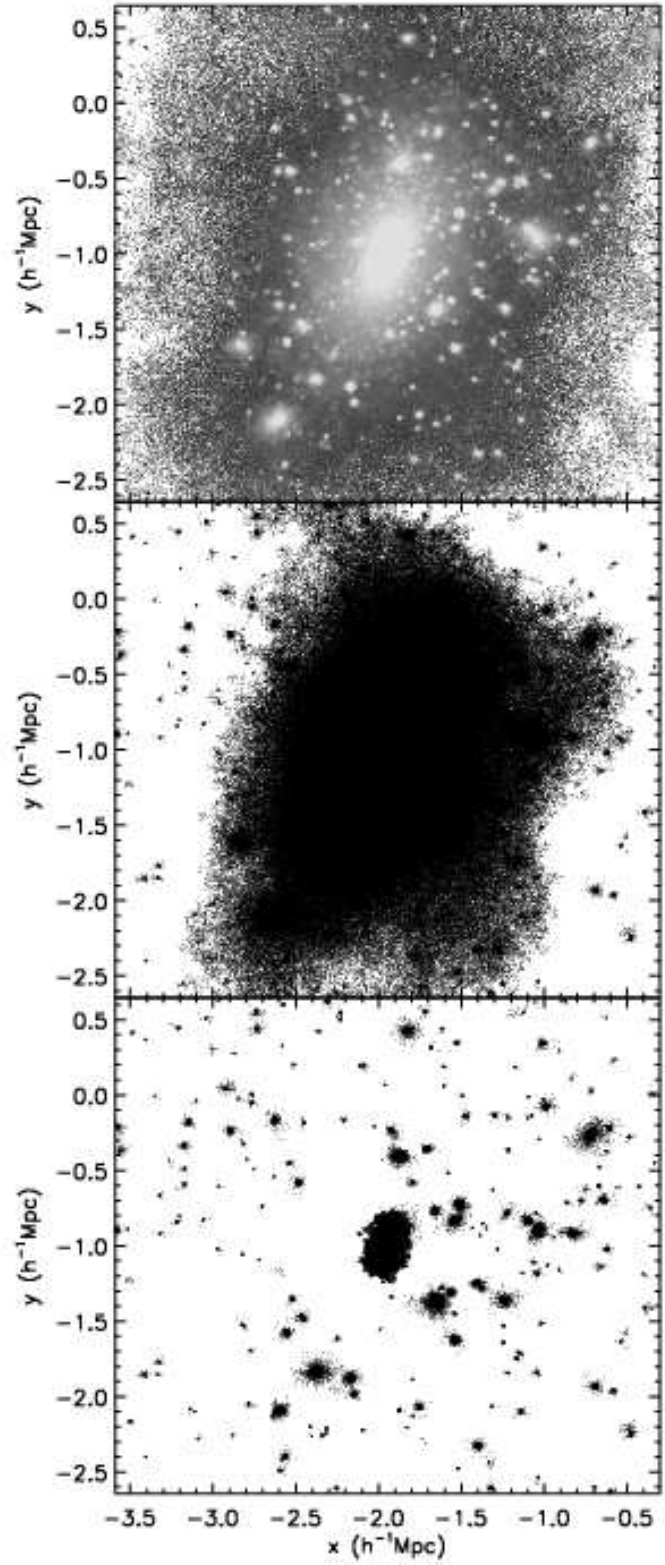}
\caption{
Density map ({\it top panel}), the $x$--$y$ projection of 
the particles contained in FOF-halos ({\it middle panel}), 
and projection of those in SKID-halos ({\it bottom panel}) 
within cluster's virial radius at $z = 0$. 
\label{fig1}}
\end{figure}

Next, we divide these FOF-halos into substructures according to their
density peaks. For this purpose, we use SKID algorithm \citep{gov97} as
a substructure finder. The particles contained in the FOF-halos are
moved along the density gradients to the local density maxima. At each
density peak, the localized particles are grouped. After we remove the
gravitationally unbound particles in each group, we identify the groups
that contain more than 10 particles as substructures.  We call these
groups {\it SKID-halos}.  The SKID-halo that contains the most bound
particle of its host FOF-halo is defined as a central SKID-halo and its
properties are replaced by those of the host FOF-halo.  In
Fig.~\ref{fig1}, we show the density map of the cluster at $z = 0$ and
the particles contained in the FOF-halos and the SKID-halos.

The merger tree of each SKID-halo is constructed by the method described
by \citet{oka00}. Note that we trace halo-stripped galaxies as well as
the SKID-halos using the most bound particle in a SKID-halo as a tracer
particle (see Okamoto \& Habe 1999, 2000).

Since cooling is not allowed in substructure halos 
(i.e. non-central SKID-halos) in our model as we will describe in 
\S \ref{hotstrip}, the non-central SKID-halos that do not have any 
progenitors at previous time-step cannot contain either stellar or 
cold gas components. 
Thus, we remove these useless SKID-halos from our merger trees. 
As a result, a newly-formed SKID-halo, i.e. a SKID-halo that has no progenitor 
at previous time-step, is always born as a central SKID-halo.

\subsection{Galaxy Formation Model} \label{SAmodel}

The following prescriptions are almost the same as the model in
\citet{on01}.

For simplicity, a dark halo is modeled as a truncated isothermal sphere
whose mass and circular velocity are taken from the $N$-body data.  If a
dark matter halo has no progenitor halos, the mass fraction of the gas
is given by $f_{\rm b} = \Omega_{\rm b}/\Omega_0 \simeq 0.1$.  
When a dark matter halo collapses, 
the gas in the halo is shock-heated to the virial
temperature of the halo.  We refer to this heated gas as {\it hot gas}.  
The gas in dense region of the halo is cooled by efficient
radiative cooling.  We dub this cooled gas {\it cold gas}.  We estimate
the amount of the gas cooled during a time-step using the
metallicity-dependent cooling function by \citet{sut93}.  
In order to avoid the formation of unphysically large galaxies, 
the above cooling process is applied only to halos with circular 
velocities smaller than $V_{\rm stop}$. 
When we adopt the same value of $V_{\rm stop}$ for the field and 
cluster simulations, our simulations produce too many/few bright galaxies 
in the field/cluster simulation. 
Although this discrepancy is interesting, our simulations are too small to 
argue the bright end of luminosity functions and it is beyond the scope of 
this paper. 
Moreover the physical meaning of $V_{\rm stop}$ is quite unclear. 
Therefore we choose the values of $V_{\rm stop}$ to reproduce the observed 
number of bright galaxies ($L > L_*$). 
We adopt $V_{\rm stop} = 350$ km s$^{-1}$ and 500 km s$^{-1}$ for the filed 
and cluster simulations, respectively.

Stars are formed from the cold gas at a rate of 
$\dot{M}_*=M_{\rm cold}/\tau_*$, where $M_{\rm cold}$ is the 
mass of the cold gas and $\tau_*$ is the star formation time-scale 
in the disk. 
It means that the cold gas in the disk is consumed in about $\tau_*$ 
if there is not any supply of the fresh gas.

When the star formation time-scale is a function of redshift as in the
case of $\tau_* \propto \tau_{\rm dyn}$, where $\tau_{\rm dyn}$ is the
dynamical time of a galaxy, the observed properties of the high redshift
galaxies cannot be reproduced \citep{kh00, som01, nag01}.  We then adopt
the description by \citet{col94},
\begin{equation}
\tau_* = \tau_*^0 \left(\frac{V_c}{300 \ {\rm km} \ 
{\rm s^{-1}}} \right)^{\alpha_*}, 
\label{sft}
\end{equation}
where $\tau_*^0$ and $\alpha_*$ are free parameters.  These free
parameters are fixed by matching the observed mass fraction of the cold
gas in neutral form in the disks of spiral galaxies.

In our model, stars with masses larger than 10 $M_{\odot}$ explode as
Type II supernovae (SNe) and heat up the surrounding cold gas.  This SN
feedback reheats the cold gas to the hot phase at a rate of $\dot{M}_{\rm
reheat} = M_{\rm cold}/\tau_{\rm reheat}$, where the time-scale of
reheating is given by
\begin{equation}
\tau_{\rm reheat} = \left(V_c/V_{\rm hot}\right)^{\alpha_{\rm hot}} 
\tau_*. 
\label{feedback}
\end{equation}
where $V_{\rm hot}$ and $\alpha_{\rm hot}$ are free parameters.
Although they are to be determined by matching the local luminosity
function of galaxies, we set $\alpha_{\rm hot}$ to 2.0 because changing
this value has a minimal effect on the results if
we determine other parameters to reproduce observations of local
galaxies.  

Chemical evolution is treated in almost the same way as described 
in \citet{kc}. The instantaneous recycling approximation is adopted. 
We assume that the initially gas has a primordial composition 
(zero metallicity). Then the gas is polluted by TypeII SNe. 
The amount of metals ejected from SNe is characterized by $y$, which 
is heavy element yield for each generation of stars.  
The value is defined by matching the color-magnitude relation 
(CMR) of the local cluster ellipticals and we adopt $y = 3 Z_{\odot}$ 
in this paper. All the ejecta is incorporated with the cold gas, while 
30 \% of the ejecta was assumed to be directly expelled to the hot gas in
\citet{kc}.  
The gas fraction returned by evolved stars is 0.25 in this paper. 
Simultaneously, the SNe heat up the surrounding cold gas, 
and then metals contained in the cold gas are also returned to the hot gas.  
The returned metals affects cooling rate of the hot gas and the cooling 
transfers the metals from the hot gas to the cold gas. 

Before we describe our merging procedure, let us explain our galaxy
hierarchy. As we mentioned before, FOF-halos represent virialized dark halos. 
Each FOF-halo hosts one central SKID-halo along with a number of
non-central SKID-halos that represent substructures of the FOF-halo. 
Each SKID-halo hosts one central galaxy along with a number of satellite
galaxies. As we will dscribe \S~\ref{hotstrip}, only the central galaxy
in the central SKID-halo can be supplied with the fresh cold gas by
radiative cooling. 
The merging between SKID-halos are directly obtained from $N$-body
simulations. We here consider the merging between galaxies within a
SKID-halo. 

When two or more SKID-halos have merged, we identify a central galaxy of
the largest progenitor as a central galaxy of the new common SKID-halo.
Other galaxies are regarded as satellite galaxies. 
For simplicity, we only take account of the merging between a satellite
galaxy and the central galaxy. 
A satellite galaxy falls into the central galaxy in dynamical friction 
time-scale,
\begin{equation}
\tau_{\rm mrg} = \frac{1}{2}\frac{f(\epsilon)}{C}
	\frac{V_c r_{\rm halo}^2}{GM_{\rm sat}\ln \Lambda_{\rm C}}, 
\end{equation}
where $r_{\rm halo}$ and $V_c$ are the radius and circular velocity of
the new common halo, respectively, $M_{\rm sat}$ is the mass of
the satellite, and $\ln \Lambda_{\rm C}$ is the Coulomb logarithm.  
If the satellite has a halo at previous time-step, 
$M_{\rm sat}$ is defined as the mass of the SKID-halo to which 
the satellite belongs as the central galaxy at the previous time-step. 
Since we trace halo-stripped galaxies as well, the mass composed by 
stars and cold gas is used as $M_{\rm sat}$ when the satellite does not 
have a halo at previous time-step. 
The function $f(\epsilon)$ describes the dependence
of the orbital decay on the eccentricity of the orbit, expressed in
terms of $\epsilon = J/J_c(E)$, where $J_c(E)$ is the angular momentum
of a circular orbit with the same energy as the satellite.  This
function is well approximated by $f(\epsilon) \simeq \epsilon^{0.78}$,
for $\epsilon > 0.02$ \citep{lac93}.  $C$ is a constant with value
$\simeq 0.43$. For $f(\epsilon)$, we adopt the average value, $\langle
f(\epsilon)\rangle \simeq 0.5$, computed by \citet{tor97}.

When a satellite galaxy merges with a central galaxy and the mass ratio
of the satellite to the central galaxy is larger than $f_{\rm major}$,
we regard this merger as a major merger, and then all stars of the
satellite and disk stars of the central galaxy are incorporated with the
bulge of the central galaxy. At the same time, cold gas contained in
both galaxies is consumed to form bulge stars by starburst with the same
feedback law as in the disk star formation.

If the mass ratio does not exceed the value of the parameter, $f_{\rm
major}$, the merger is classified as minor.  When we adopt the standard
prescription, in the case of the minor merger, the disk of the central
galaxy is preserved, no starburst takes place, and the stars from the
satellites are added to the bulge of the central galaxy. 
%
%
In this paper we also consider the case that a minor merger triggers
small starburst because numerical experiments of merging galaxies have
suggested that starburst is induced by a minor merger as well as a major
merger \citep{wal96}. We dub this process {\it minor burst} and will
give the detailed prescription in \S \ref{minib}.

In this paper, we adopt $f_{\rm major} = 0.2$ that was used for the
low density universe in \citet{on01}. We confirmed that our results
hardly depend on $f_{\rm major}$, since our morphological classification
based on $B/T$ is normalized to reproduce the
observed morphological composition in the field.

The absolute luminosities and colors of individual galaxies are
calculated using a population synthesis model by \citet{kod97}.  
The initial stellar mass function (IMF) that we adopt is the power-law 
IMF of Salpeter form with lower and upper mass limits of 0.1 and 60 
$M_{\odot}$, respectively.
 
As we stated above, we divided the stellar component into disk and bulge
components.  In the previous SA studies, morphology of each galaxies
have been determined by the $B$-band $B/T$. \citet{sim86} showed that 
the Hubble types $T$ of galaxies correlated with $B/T$ by following relation,
\begin{equation}
<\Delta M> = 0.324(T + 5) - 0.054 (T + 5)^2 + 0.0047(T +5)^3,
\label{morphology}
\end{equation}
where $\Delta M \equiv M_{B}^{\rm bulge}-M_{B}^{\rm total}$.  In this
paper we use this relation to assign the Hubble type to our model
galaxies. Unfortunately this relation has significant scatter
\citep{bau96}, therefore we prefer direct comparison of $B/T$s of the
model galaxies with those of the observed galaxies as we will describe
in \S \ref{morph-grad}.

\subsubsection{Starburst triggered by a minor merger} \label{minib}

The effect of minor mergers was studied in \citet{mh94} using SPH
simulations including star formation.  In the case of a merger between a
Milky Way-sized disk galaxy and a satellite with total mass equal to
10\% of the central disk galaxy mass, the bar instability leads a gas
fueling to the galactic center, and then 50 \% of the total gas is
consumed in the starburst.  If the central galaxy has a bulge, the bulge
stabilizes the disk and prevents the gas fueling.  When the
bulge-to-disk mass ratio is larger than 1/3, the starburst becomes
negligible.

\citet{som01} modeled this process by assuming that the fraction
of cold gas consumed by the starburst is a function of the merging mass
ratio.  However, since only limited parameter range has been explored,
how the starburst efficiency depends on the mass ratio is highly
unknown.  Moreover, the starburst during a violent merger and the
starburst triggered by the bar instability should be treated separate
way.  Then, we model the minor burst by assuming that the fraction,
$f_{\rm burst}$, of the cold gas is consumed by the minor burst when the
merger mass ratio is between $f_{\rm minor}$ and $f_{\rm major}$ for
simplicity.  In the case of the minor burst, the stellar disk of the central
galaxy is preserved.  When the bulge-to-disk mass ratio of the central
galaxy is larger than 1/3, the minor burst does not take place according to
\citet{mh94}.  We set these two parameters to reproduce the
morphology-radius relation ($f_{\rm burst} = 0.25$ and $f_{\rm minor} =
0.1 f_{\rm major}$), as we will show in \S \ref{morph-grad}.

Note that if we adopt $f_{\rm burst} = 1$, it leads very similar results
to the model without the minor burst and with small $f_{\rm major}$ ($\simeq
f_{\rm minor}$) in clusters, because most of gas is consumed by the
mergers at high redshifts in any case.  The model with small $f_{\rm
major}$ only increases the bulge-dominated fraction and does not
increase the intermediate $B/T$ fraction \citep{on01}.

\subsubsection{Stripping of Diffuse Hot Gas} \label{hotstrip}

We include the stripping of diffuse hot gas from galactic halos in our
modeling as follows.  When a FOF-halo contains more than two SKID-halos,
we identify the central SKID-halo and properties of this central
SKID-halo are replaced with those of the host FOF-halo as we mentioned
in \S \ref{mergertree}.  The hot gas contained in the non-central
SKID-halos are simply added to the hot gas reservoir of the host
FOF-halo (i.e. the central SKID-halo), and so, only the central galaxy
in the central SKID-halo can be supplied with the fresh cold gas by
radiative cooling.  This process has already been included in the
previous SA studies and led significant success
\citep[e.g.][]{bau96,kc,dia01}.

\subsubsection{Ram pressure Stripping of Cold Gas Disk}

The RPS of cold disk gas is also a new ingredient of our model.  By
following simple method, we estimate stripping effects of cold gas by RP
from an ICM.

In order to estimate the strength of the RP, we need to know orbits of
cluster galaxies. For this purpose, we assume that the cluster has a
NFW-type of density distribution \citep[][]{nfw97}, and then
we fit the density profile of the main cluster (i.e. the most massive
virialized object) at each time-step of the merging history by the NFW
profile,
\begin{equation}
\rho(r) = \frac{\rho_0}{(r/r_s)(1+ r/r_s)^2},
\label{nfw}
\end{equation}
where $\rho_0$ and $r_s$ are fitting parameters.  Once we fit the
density profile of a cluster, we can calculate the potential energy as a
function of the distance from the cluster center,
\begin{equation} 
\Psi(r) = -4 \pi r_s^2 G \rho_0 \frac{\ln(1 + r/r_s)}{r/r_s} 
	- \frac{1}{2} H_0^2 \lambda_0 r^2.
\label{pot}
\end{equation}
Then we calculate the orbits of the galaxies within $2R_{\rm vir}$ from 
the center of the main cluster using their positions and relative
velocities to the cluster.

Although gradual removal is preferable, we here consider only
instantaneous stripping as a maximal effect because we have to output
data with a quite small time-step to achieve the gradual removal of gas
and such a treatment is not practical.  Thus, the RP is estimated at
pericentric positions for outgoing galaxies and at present positions for
ingoing galaxies.  This treatment avoids too early stripping for the
ingoing galaxies that do not pass the pericentric positions during the
time-step.

The RP from the ICM is estimated as follows at the stripping positions,
\begin{equation}
P_{\rm ram} = \rho_{\rm ICM} v_{\rm gal}^2, \label{rp}
\end{equation}
where $\rho_{\rm ICM}$ is the density of the ICM and $v_{\rm gal}$ is 
the velocity of a galaxy relative to the ICM. 
We suppose that the ICM distributes parallel to the dark matter for simplicity.  
The density of the ICM is obtained as $f_{\rm b}\rho(r)$, where $\rho(r)$ 
is taken from the NFW fit.  
The hot gas fraction in the main halo at $z < 1$ is $\sim 0.75$, however 
using $0.75 f_{\rm b} \rho(r)$ as $\rho_{\rm ICM}$ makes no visible 
difference. Thus, we assume the hot gas fraction is unity for simplicity.

We also attempted the ICM distribution that has the Coma-like core with the 
core density,  $\rho_c = 7.98 \times 10^{-27} h^{\frac{1}{2}}$ g cm$^{-3}$ 
\citep{bri92} by putting the ceiling to the density distribution. 
With the above choice of the core density, the core radius becomes $r_{\rm c} 
\simeq 140 h^{-1}$ kpc ($\sim 0.12 R_{200}$) at $z = 0.2$. 
This radius is much smaller than the radius at which the effect of the RPS 
becomes significant as we will show in \S \ref{SFRG}, and then we cannot 
find any difference in the results from the coreless distribution.  
Therefore we show only the results in the case of the ICM distribution 
parallelized to dark matter.  

For the cold gas, the restoring force per unit area due to the gravity
of the galactic disk is given by

\begin{equation}
F_{\rm grav, cold} = 2 \pi G \Sigma_{*,{\rm disk}} \Sigma_{\rm cold},
\label{grav}
\end{equation}
where $\Sigma_{*,{\rm disk}}$ is surface density of the stellar disk and
$\Sigma_{\rm cold}$ is that of the cold gas \citep{gun72}.  We calculate
$\Sigma_{*,{\rm disk}}$ as $0.5 M_{*,{\rm disk}}/\pi r_{\rm e}^2$, where
$r_{\rm e}$ is an effective radius of the disk assuming that the
effective radius is identical to the half-mass radius of the disk. 
The effective radius is given by matching the observed luminosity-radius 
relation of spiral galaxies (see \S \ref{param}).  
Assuming $\Sigma_{\rm cold} \propto \Sigma_{*,{\rm disk}}$, 
the surface density of the cold gas is given
by $\Sigma_{\rm cold} = (M_{\rm cold}/M_{*,{\rm disk}}) \Sigma_{*,{\rm
disk}}$.

Since we consider the maximal effect of the RPS, we suppose that all cold
gas in a disk is stripped when $P_{\rm ram} > F_{\rm grav}$.  The
stripped cold gas is mixed with the hot gas of the host FOF-halo.

\section{Model Sets}

In this paper we investigate the following four baseline models
and their combinations to clarify effect of each process.

As the standard prescription of SA, we consider the model without either
the RPS or the minor burst. By this model we define a reference
parameter so that this model reproduces the properties of the local galaxies. 
The model with the reference parameter set is called the {\it standard}
model. 
When no fresh cold gas is supplied, the cold gas in a galaxy is exhausted by 
the star formation and then the star formation is truncated. 
Since this truncation time-scale is determined by the star formation  
time-scale, $\tau_*$, we also consider the model with longer star formation
time-scale than the standard model. 
The model with the star formation time-scale that is four times as long
as the standard one is referred as the {\it low SFR} model.  
The model with the minor burst and the model with the RPS are called the
{\it minor burst} model and the {\it RPS} model, respectively. In these
models, the reference parameter set is used.  

All models are studied in both the cluster and field simulations except
for the RPS model that is unimportant in the field. Thus, the
corresponding model to the cluster standard and the RPS models in the
field is the field standard model.  
When we compare the cluster simulation to the field simulation, 
we always use the corresponding field model, for example the
minor burst model in the field for the minor burst model in the cluster
and the low SFR model in the field for the low SFR + RPS model in the
cluster. These models are simply dubbed the {\it field} models.

\subsection{Parameter setting} \label{param}

In this subsection we show characteristics of our models and how each
process affects the properties of the local field galaxies. The values
of the parameters used in the standard model are listed in Table~\ref{tbl-2}.

\begin{figure}
\epsscale{0.8}
\plotone{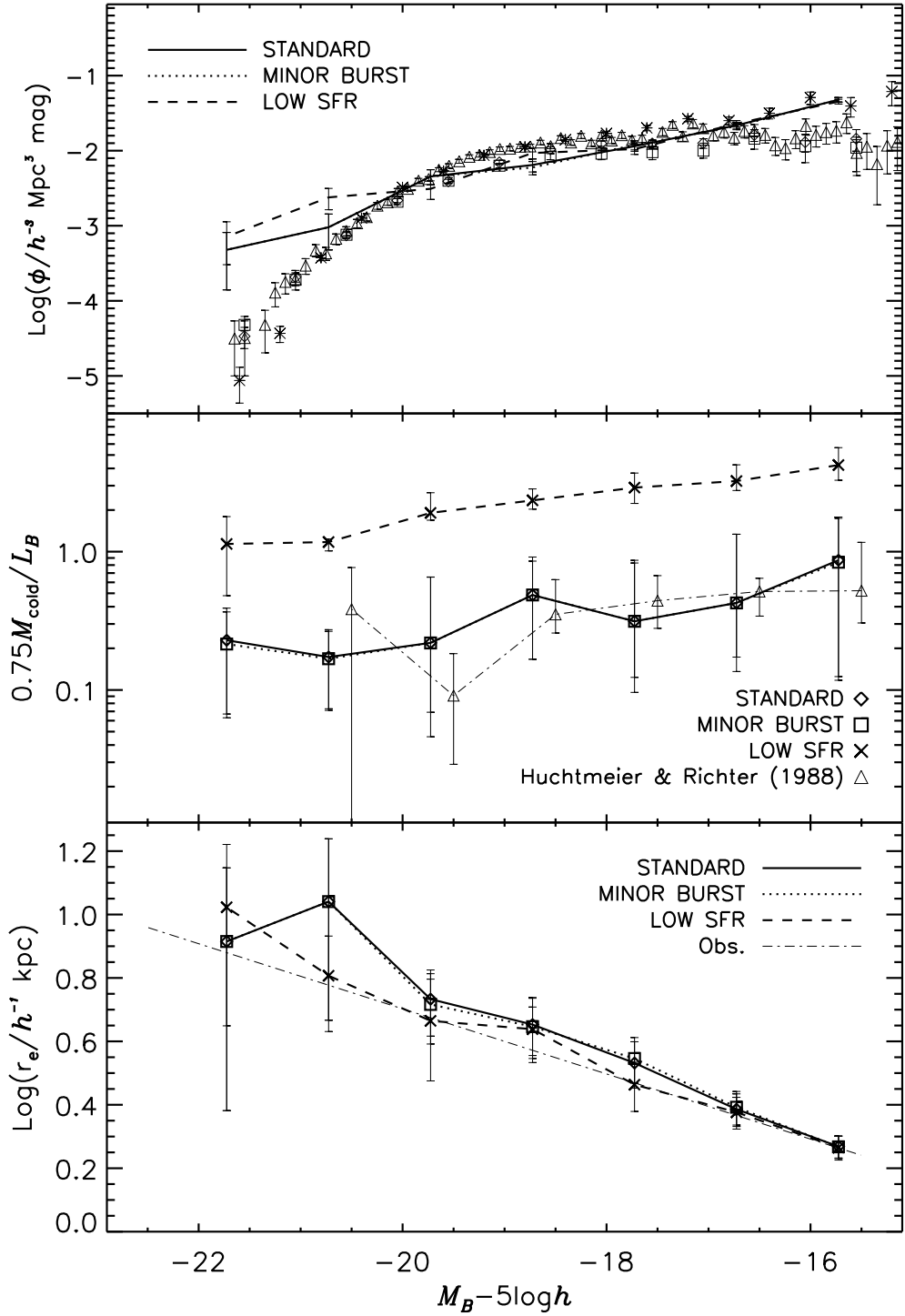}
\caption{
Properties of the galaxies in the field models as a function of $B$-band luminosity. 
$B$-band luminosity functions, cold gas mass fractions in spiral galaxies, and size of spiral galaxies are presented in the top, middle, and bottom panels, respectively. 
The solid, dotted, and dashed lines indicate the standard, minor burst, and low SFR models, respectively. 
The observed data are represented by symbols for the luminosity functions (squares, asterisks, diamonds, and triangles for APM, ESP, UKST, and 2dF, respectively) and dash-doted lines for the cold gas mass fractions and the galaxy size. The details are given in the text. 
\label{fig2}}
\end{figure}

In the top panel of Fig.~\ref{fig2} we show the model $B$-band
luminosity functions in the field at $z = 0$ and the observed one taken
from APM \citep{lov92}, ESP \citep{zuc97}, UKST \citep{rat98}, and 2dF
\citep{fol99}.  
Since the simulation volume for the field is too small, 
the brightest bin in the standard model contains only two
galaxies. Therefore we do not attempt to reproduce the bright 
end of the field luminosity function in this paper. 
Apart from the bright end, our luminosity functions are smaller 
at the intermediate luminosity range than those in the observations 
and their slopes at faint-ends are steeper than observed slopes. 
This difference may cancel out if random collisions between satellite 
galaxies (in substructure halos) are considered \citep{som99} or 
if we use higher resolution that enables us to identify dark halos of 
smaller satellite galaxies \citep{spr01}.

It is found that the minor burst hardly affects the galaxy
luminosity function at least in the field.  The luminosity functions in the
standard and minor burst models almost overlap.  The effect of the star
formation time-scale is also small as shown in \cite{col00}.  

Although a model luminosity function is very sensitive to the
choice of the feedback parameters ($V_{\rm hot}, \alpha_{\rm hot}$) 
and the circular velocity above which the cooling is stopped 
($V_{\rm stop}$), we note that the shape of the luminosity function does
not have particular importance for our results. 
The key observational constraints in this paper are the number of bright
galaxies ($L > L_*$), the cold gas mass fraction in a late type galaxy,
and the disk size. The number of bright galaxies are controlled by
$V_{\rm stop}$, and latter two are used to specify our model parameters
as described below. 

When examining the effects of gas stripping, one of the most important
quantities is the cold gas fraction in galaxies that reside in the
environment where the stripping effects are negligible.  The cold gas
content of galaxies are strongly affected by the star formation
parameters, $\tau_*^0$ and $\alpha_*$ \citep{col00}.  In the middle
panel of Fig.~\ref{fig2} we show the cold hydrogen fraction of the
spiral and irregulars in the field models. The observational data are
taken from the \ion{H}{1} observation by \citet{hr88}, which come from a
complete sample of the galaxies with morphological type $T \geq 1$.  We
classify our galaxy according to Eq. (\ref{morphology}) and choose the
galaxy with $T \geq 1$.  It is found that the models with the reference
parameter set reproduce the observation well. There is no difference
between the standard model and the minor burst model again.  The
fraction in the low SFR model is much larger than that in the standard
model and the observation as expected.

Another important quantity when we argue the RPS is the size of a disk that
is used to calculate the surface density of the disk. We assume that the
disk size is proportional to the virial radius of its host halo. In the
bottom panel of Fig.~\ref{fig2} we show the case that $r_{\rm e}$ is 5
\% of the virial radius.  We also show the observed best-fit relation
given by \citet{ty00} for the data in \citet{imp96}.  It is found
that our model gives a good agreement with the observed relation.

One important conclusion in this subsection is that the minor burst hardly
affects the properties of the field galaxies with our values of 
$f_{\rm minor}$ and $f_{\rm burst}$ at least at $z = 0$. 

\section{RESULTS}

\subsection{Global Properties}

Before we investigate the distribution of galaxy population, it is
worthwhile looking over global properties of our cluster galaxies.

\begin{figure}
\epsscale{1.0}
\plotone{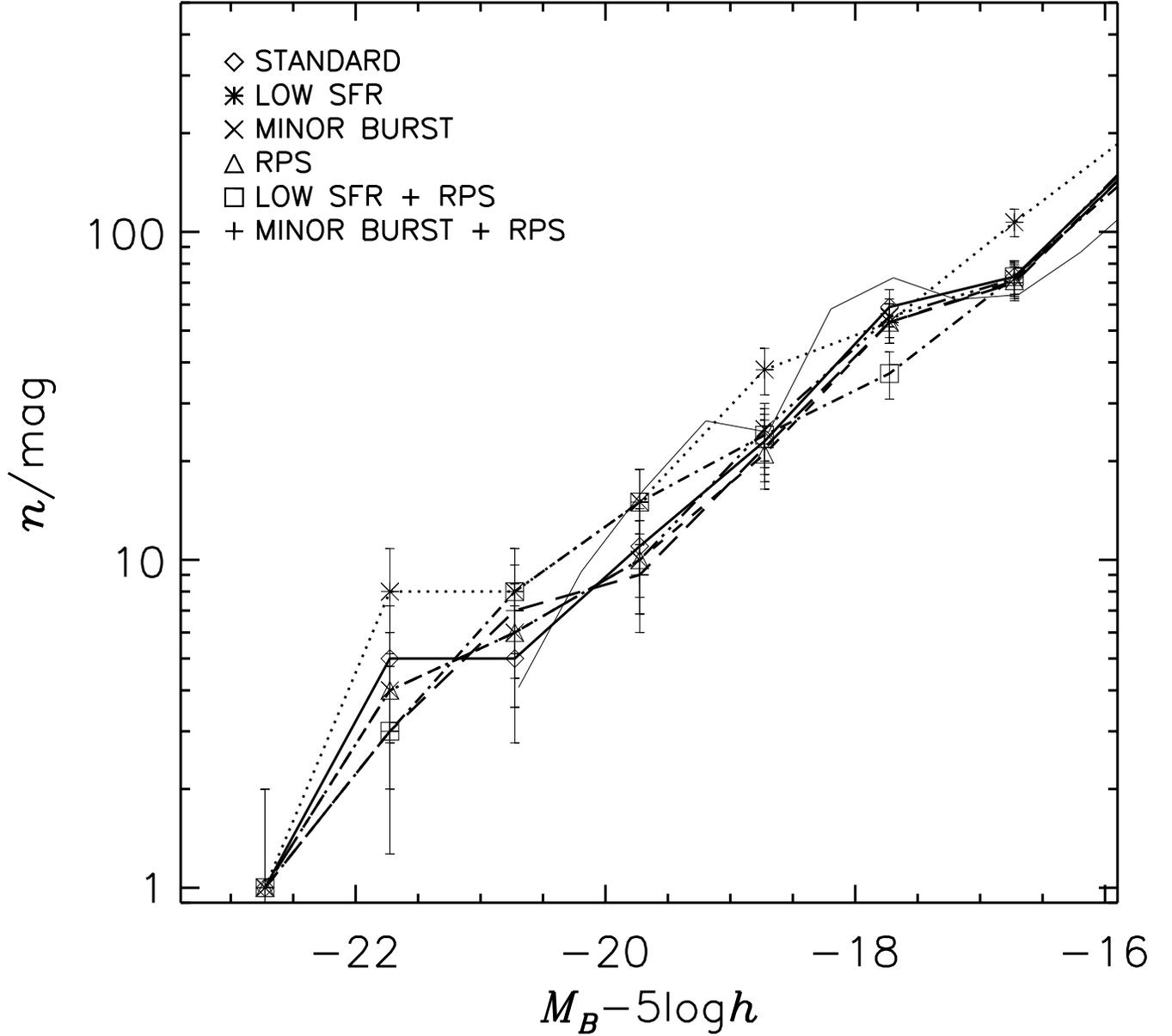}
\caption{
$B$-band luminosity functions of cluster galaxies. 
The diamonds, asterisks, crosses, triangles, squares, and pluses indicate the luminosity functions of the standard, low SFR, minor burst, RPS, low SFR + RPS, and minor burst + RPS models, respectively. 
The thin solid line is the luminosity function of the Virgo cluster galaxies by \citet{san85}. 
\label{fig3}}
\end{figure}

In Fig.~\ref{fig3}, we show $B$-band luminosity functions of cluster
galaxies in the standard, low SFR, and minor burst models and their
combination with the RPS at $z = 0$.  We also show the observed luminosity
function in the Virgo cluster \citep{san85}. All models broadly
agree with the observation. The luminosity function in the low SFR
model is always larger than those in other models because $B$-band
luminosity reflects recent star formation.  
The RPS affects only the low SFR model significantly, especially at the 
faint end, because the galaxies in this model have more cold gas than
those in the models with the standard star formation time-scale and the 
cold gas in the faint galaxies are easily stripped. 

\begin{figure}
\epsscale{0.8}
\plotone{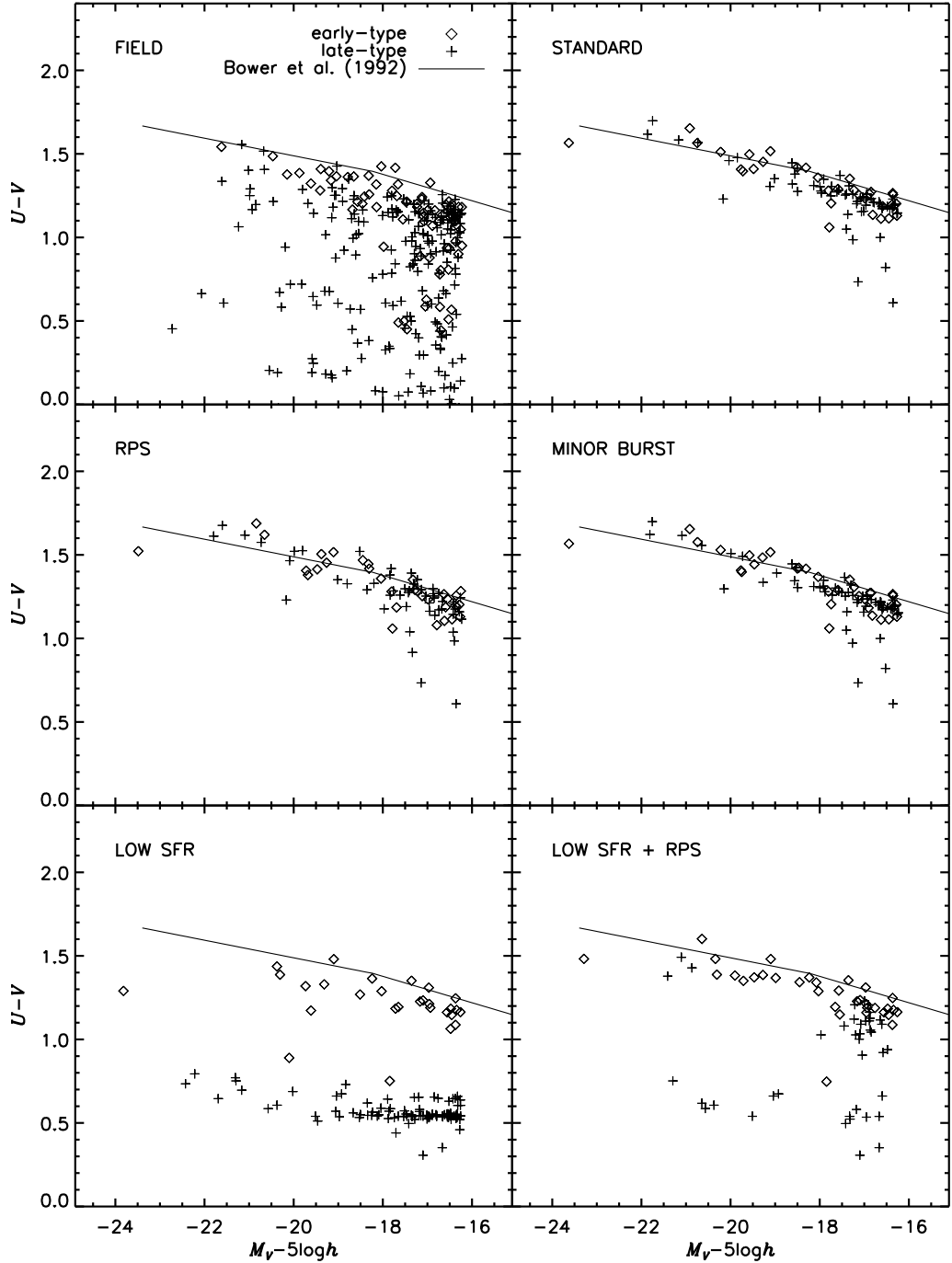}
\caption{
Color-magnitude diagrams at $z = 0$. 
We show $U-V$ colors of galaxies in the field model ({\it top left}) and galaxies within the cluster core in the standard ({\it top right}), RPS ({\it middle left}), minor burst ({\it middle right}), low SFR ({\it bottom left}), and low SFR + RPS ({\it bottom right}) models as a function of $V$-band luminosity. 
Galaxies with $T < 0.92$ by Eq. (\ref{morphology}) are classified as early-type galaxies (diamonds) and other galaxies are classified as late-type galaxies (pluses). 
The solid lines show the observed color-magnitude relation for cluster ellipticals by \citet{bow92}, which are applied aperture correction \citep{kod98}. 
\label{fig4}}
\end{figure}

To see effects of truncation of star formation on colors of galaxies,
in Fig.~\ref{fig4} we show CMRs for all types of galaxies in 
the standard field model and galaxies within cluster cores 
($r < 0.5 h^{-1}$ Mpc) in the standard, RPS, minor burst, 
low SFR and low SFR + RPS models.  
We do not show the minor burst + RPS model because it is quite similar to
the RPS model.  Here we classify the galaxies with $T < 0.92$ according
to Eq. (\ref{morphology}) as early-type galaxies (diamonds) and others
as late-type galaxies (pluses).  The observed CMR for the Coma
ellipticals is indicated by the solid lines \citep[][]{bow92}, which is
applied aperture correction \citep{kod98}.  In the models with the
proper star formation time-scale (standard, RPS, and minor burst), the
blue population seen in the field model is vanished, and then the
galaxies distribute within $\pm2$ mag from the CMR for the observed
cluster ellipticals as observed in the Coma cluster \citep{ter01}.  This
is caused by the hot gas stripping in the standard model. The earlier
formation epochs of galactic halos in the cluster environment may not be
important for this because the late-type galaxies in the low SFR model
have significantly bluer colors than the standard model.  On the other
hand, the CMRs for early-type galaxies are quite similar in all models,
even in the field. It suggests that the main star formation truncation
mechanism for the early-type galaxies is the starburst caused by 
the major merger.

Since the star formation is sufficiently suppressed in the standard
model, the significant difference by the RPS is only seen between the low
SFR model and the low SFR + RPS model. However, even in the low SFR +
RPS model, there are a lot of too blue galaxies. Thus we conclude that
the most important star formation truncation mechanism for the
disk-dominated galaxies in cluster cores is the hot gas stripping.  
We cannot see any significant effect by the minor burst on the CMR.

\subsection{SFR Gradients} \label{SFRG}

Recently, many authors have analyzed radial trends in the SFR
\citep{bal98, bal99, ell01} of galaxies in the CNOC1 sample
\citep{yee96} .  Here we investigate what mechanism is responsible for
the SFR gradient with the special interests in the effect of the RPS.

\begin{figure}
\epsscale{0.8}
\plotone{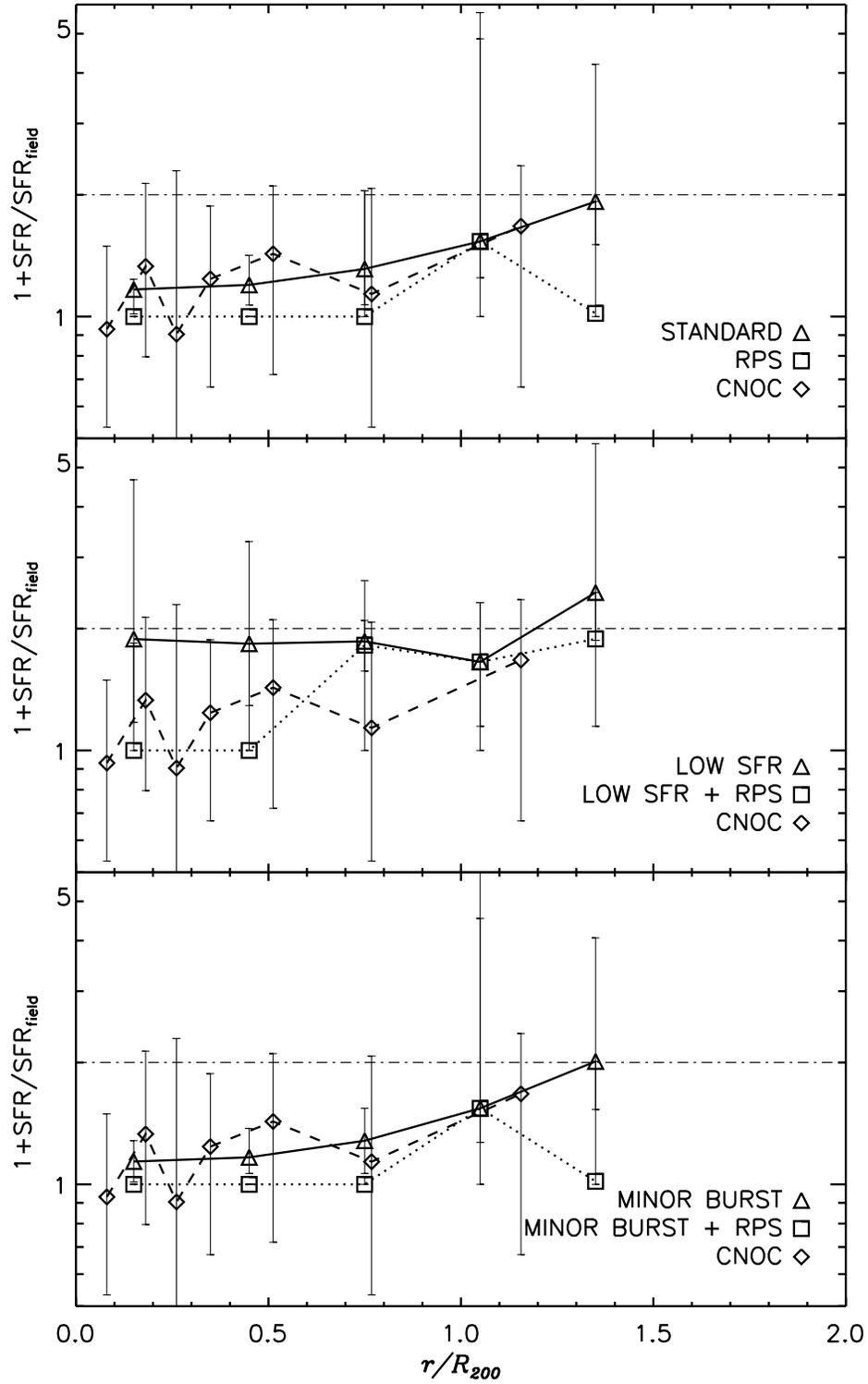}
\caption{
The median SFRs of galaxies with $M_R < -20.5$ are plotted as a function of projected radius at $z = 0.2$. 
Solid lines in the top, middle, and bottom panels show the SFRs in the standard, low SFR, and minor burst models, respectively. The corresponding RPS models are indicated by the dotted lines. 
The observed median SFR of the galaxies in the CNOC sample is taken from \citet{dia01} and shown by the asterisks. 
The dot-dashed line indicates the median SFR in the field model. 
An error bar shows the 25th to 75th percentile of the distribution in each bin.
\label{fig5}}
\end{figure}

In Fig.~\ref{fig5} we show median SFRs for observations and our models
as a function of projected radius. The observed SFR distribution taken
from \citet{dia01} is indicated by diamonds. The galaxies brighter than
$M_R = -20.5$ are selected assuming the same cosmology as our
simulations.  In this figure we show the low redshift sample of
\citet{dia01}. The data were constructed by superposing 7 CNOC1
clusters at the redshift range between $0.18 < z < 0.3$.  Hereafter, we
refer this superposing sample as the {\it CNOC sample} in this paper.
The details are given in \citet{dia01} and references therein.
We show relative SFRs, which are normalized by the field median value 
obtained from the same redshift range, because the observed SFRs
determined by $W_{O_{\rm II}}$ have a lot of uncertainties.  
Both for the observational and simulation data, we exclude the brightest 
cluster galaxies because they often have quite different properties from 
other cluster galaxies.  
The standard, low SFR, and minor burst models are represented in the top,
middle, and bottom panels, respectively. Corresponding RPS models are
also shown in each panel.

Except for the low SFR model, all our models are broadly consistent with
the observation. As expected, the low SFR model has much closer SFR to
the field SFR than other models. If we adopt the star formation
time-scale that reproduces the observed amount of the cold gas in the
field galaxies, the difference by the RPS is small compared with the 
observational errors.  While it is interesting that the effect of the RPS
can be seen even at $r/R_{200} \sim 1.5$, it should not be
overinterpreted when comparing with observations because the simulated
region by the high resolution particles is spherical and then the depth of
the line of sight at outermost bin through the high resolution region is
quite thin. 
As a result, in our simulation the number of galaxies at the outer regions 
of the cluster becomes small and the field contamination is
substantially underestimated. 
For direct comparison with the observations at large radii, we need
larger simulations. 

We confirm that the minor burst hardly affects the SFR distribution.

\subsection{Color Gradients} \label{sfr-gr}

Color of a galaxy reflects SFR of the galaxy and it is a more reliable
observable than the SFR.  It has been known that galaxies in inner
regions of clusters are redder than those in outer regions \citep{bo84}.
Recent observations have confirmed the existence of the color gradients,
and besides, it continues to large radii beyond the virial radii ($ \sim
2 R_{200}$) \citep{abr96, car97}.

\begin{figure}
\epsscale{0.8}
\plotone{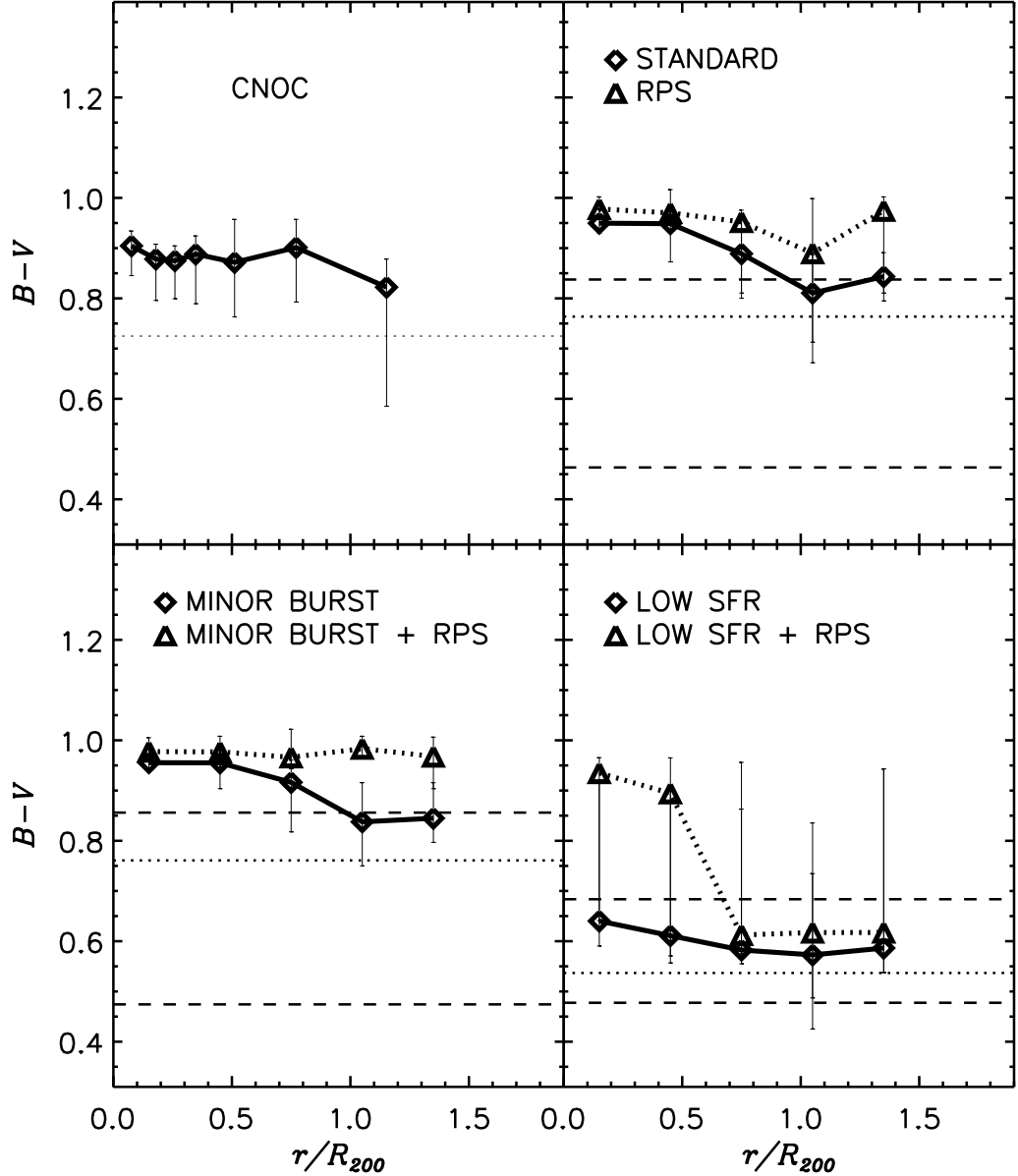}
\caption{
The median rest frame $B-V$ colors of cluster galaxies with $M_R < -20.5$ are plotted as a function of projected radius.  
In the upper left panel, the diamonds denote the observed median colors of the CNOC sample, which are taken form \citet{dia01}. 
The median color of field galaxies with $M_R < -20.5$ at the same redshift range are shown by the dotted line.
The diamonds in the upper right, lower left, and lower right panels indicate the median colors of the galaxies at $z = 0.2$ by the standard, minor burst, and low SFR models as a function of projected radius, respectively.  
The corresponding RPS model is shown by the triangles in each panel. The median color of the galaxies in the field model with $M_R < -20.5$ at $z = 0.2$ is also shown by the dotted line in each panel. 
An error bar shows the 25th to 75th percentile of the distribution in each bin. 
We also show the 25th to 75th percentile of the distribution for the galaxies in the field model by the dashed lines. \label{fig6}}
\end{figure}

In Fig.~\ref{fig6} we show the $B-V$ colors of the cluster galaxies as a
function of the projected radius at $z = 0.2$.  As the observational
counterpart, we use the same sample that we used for the SFR gradients.  
Thus, we use the galaxies brighter than $M_R = -20.5$ again.  
The median color of field galaxies is shown by dotted lines.

For the standard and minor burst models, effect of the RPS is more
significant at large radii, because, in the central part, the star
formation is strongly suppressed by the stripping of the hot gas, 
regardless of the RPS. 
Although it is difficult to say if the models with the RPS show better
agreement with the observation because our models are less contaminated
by field population as we mentioned previously, 
the standard/RPS model seems to show too steep/flat gradient compared
with the observation. It may suggest that the RPS somewhat affects the
color gradients observed in clusters. 
On the other hand, for the low SFR model, the suppression of the star 
formation by the stripping of the hot gas is quite weak and the cluster 
galaxies have similar colors to the field sample in this model. 
Then the suppression of the star formation is dominated by the RPS in 
the Low SFR + RPS model. 
It is found that the effect of the RPS is effective at $r < 0.5 R_{200}$.

While the minor burst makes galaxies slightly redder than the standard 
model, the models with and without the minor burst still produces very 
similar results.

\subsection{Morphology} \label{morph-grad}

Morphologies of cluster galaxies have a strong dependence on their
environments and redshifts. Here we investigate how each process affects
the morphologies of the model galaxies as functions of projected
radius and redshift.

\subsubsection{Morphology distribution} \label{morph-r}

Recently, some authors have studied the morphology-density or
morphology-radius relations by using hybrid method of $N$-body
simulations and SA models \citep{on01, dia01, spr01}.  
Okamoto \& Nagashima and Diaferio et al.  concluded that the morphology 
evolution only by major mergers cannot explain the S0 or intermediate 
$B/T$ population in clusters and additional processes are required 
to understand this population.  
While Springel et al. seemed to reproduce the observational
data of \citet{whi93}, they adopted wider $B/T$ range for S0's ($0.096 <
B/T < 0.45$) than the standard classification ($0.4 < B/T < 0.6$).  
The small threshold value between spirals and S0's and wider
range of $B/T$ may imply necessity of the additional processes.

One possible process is the RPS of cold gas from a galactic disk
\citep{gun72, fn99}.  According to Fujita \& Nagashima, at a central 
part of a cluster, the SFR of the disk component rapidly drops owing to 
the RPS, and then the galaxy becomes red and the disk becomes dark.  They 
suggested that a Sb galaxy is turned into a S0 galaxy as a result of the
RPS.  Another possibility is the minor merger induced bulge formation.
\citet{mh94} showed that gas supply to a galaxy center by a minor
merger triggers a star burst, and then the galaxy evolves toward the
early-type gradually.

We here examine how these processes affect the morphologies of cluster
galaxies by comparing our galaxy populations in the minor burst and RPS
models with those in the standard model and the CNOC sample.  Since we
do not have any information about galaxy morphology in our models except
for $B/T$, we classify the galaxies based on their $B/T$s  both 
in our models and the CNOC sample.  According to \citet{bal98}, we
classify galaxies into three classes: a bulge-dominated class, a
disk-dominated class, and an intermediate class.  In the CNOC
sample, \citet{dia01} have split the galaxy population based on the
$r$-band $B/T$, and then galaxies with $B/T > 0.7$ are identified as
bulge-dominated galaxies, $B/T < 0.4$ as disk-dominated galaxies, and
$0.4 < B/T < 0.7$ as intermediate galaxies.  We classify our model
galaxies based on the $B$-band $B/T$.  The boundaries between the
three classes are chosen so that the morphological fractions in
the field models match those of the field galaxies in the CNOC
sample.  As a result, the galaxies with $B$-band $B/T > 0.4$ are
identified as bulge-dominated galaxies, $B/T < 0.25$ as disk-dominated
galaxies, and $0.25 < B/T < 0.4$ as intermediate galaxies in the
standard models, and $B/T > 0.5$ for the bulge-dominated and $B/T < 0.3$
for the disk dominated galaxies in the minor burst models.

\begin{figure}
\epsscale{1.0}
\plotone{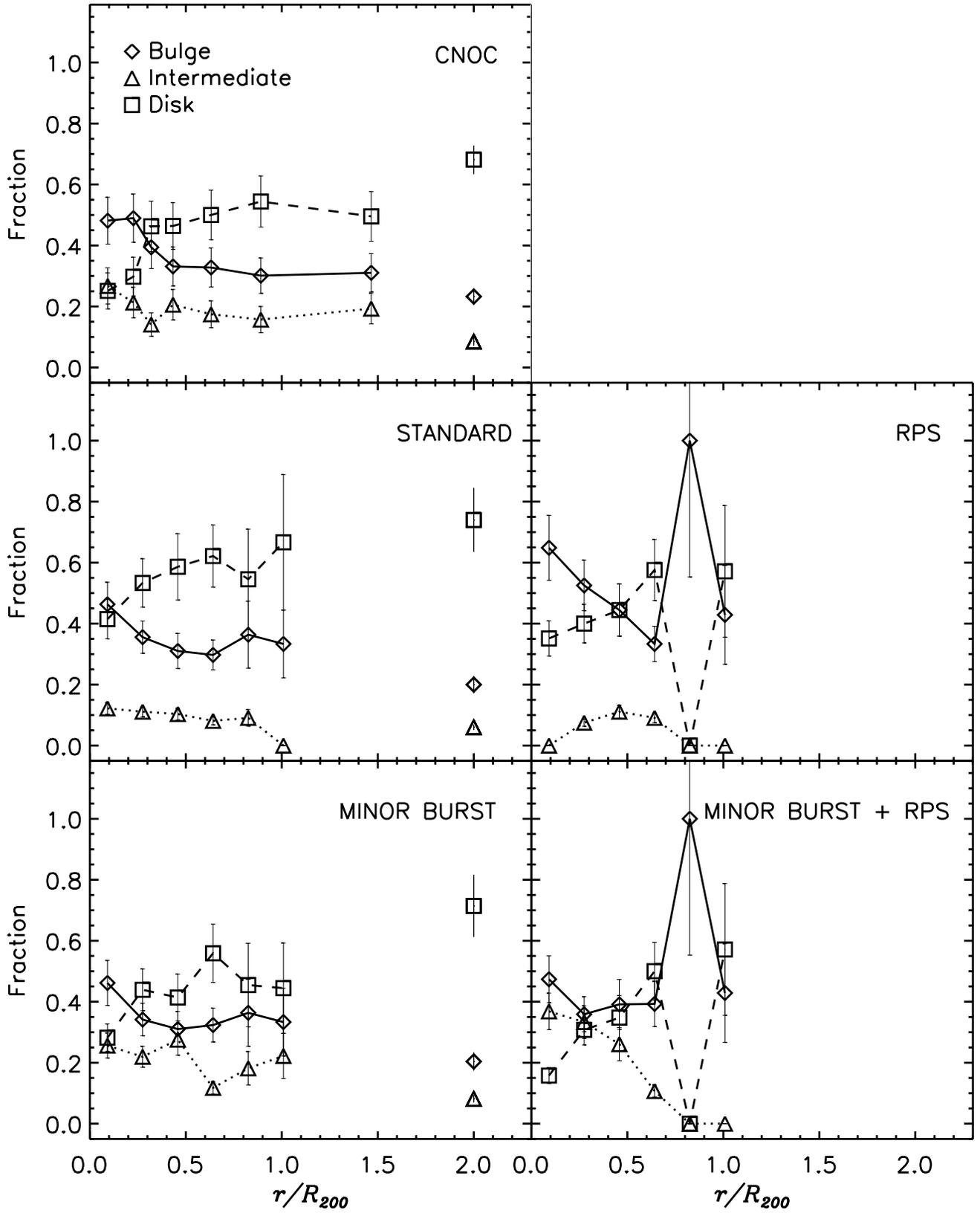}
\caption{
The morphological fractions are plotted as a function of projected radius.
The diamonds, triangles, and squares indicate the bulge-dominated, intermediate, and disc-dominated fractions, respectively. 
In the top panel we show the observed fractions for the galaxies 
with $M_R < -20.5$ in the CNOC sample, which are taken from 
\citet{dia01}.  
Other panels show the fractions in the standard, RPS, minor burst, and minor burst + RPS models for the simulated cluster at $z = 0.2$. 
The symbols at $r/R_{200} = 2$ indicate the fractions in the field in each panel. 
The error bars present the Poisson errors. 
\label{fig7}}
\end{figure}

The fraction of galaxies with $M_R < -20.5$ in each class is shown in
Fig.~\ref{fig7} as a function of projected clustercentric radius and
that of the field galaxies are shown at $r/R_{200} = 2$ in left panels.
By comparison of the CNOC sample with the standard model, it is
confirmed that the intermediate fraction in our model is much smaller 
than the observation, while the bulge-dominated fraction in our model 
agrees with the observation. 
This problem has been indicated as a shortcoming of the major merger 
induced morphology evolution model \citep{on01,dia01}.

When we consider the RPS, although the bulge-dominated fraction is increased
and the disk-dominated fraction is decreased, the intermediate fraction
is still small and it is even decreased at the center. It is because
most of the disk-dominated galaxies are almost pure disk galaxies as we will 
show and then the RPS simply darkens the disk-dominated galaxies rather 
increases their $B/T$s. 
Furthermore the intermediate galaxies with relatively large $B/T$s are 
changed into the bulge-dominated galaxies.  
The strange behavior seen in the RPS models at $r/R_{200} > 0.7$ is 
resulting from the small number of galaxies at the radius because of 
the spherical high resolution region as we discussed previously.  
At that radius, all the disk-dominated galaxies become darker than the 
luminosity cut-off by the RPS by chance. 
We expect that such a behavior are not  observed if we consider field 
contamination properly and increase the number of clusters. 
In fact, when we adopt a fainter luminosity cut-off, this behavior is 
not observed because the number of galaxies is increased.

The minor burst increases the intermediate fraction without changing the
bulge-dominated fraction because it affects only galaxies with small 
bulge-to-total mass ratios. As a result, the minor burst model
reproduces the observed fractions quite well in spite of it
does not influence other properties of galaxies.  
Now the disk-dominated galaxies tend to be non-pure disk galaxies. 
Consequently, the RPS can change their $B/T$s by the fading of their disks.  
In the minor burst + RPS model, we can see the increasing intermediate fraction
towards the center.  It indicates that the disk-dominated galaxies have
larger bulges in the inner part of the cluster resulting from the minor
burst.

\begin{figure}
\epsscale{1.0}
\plotone{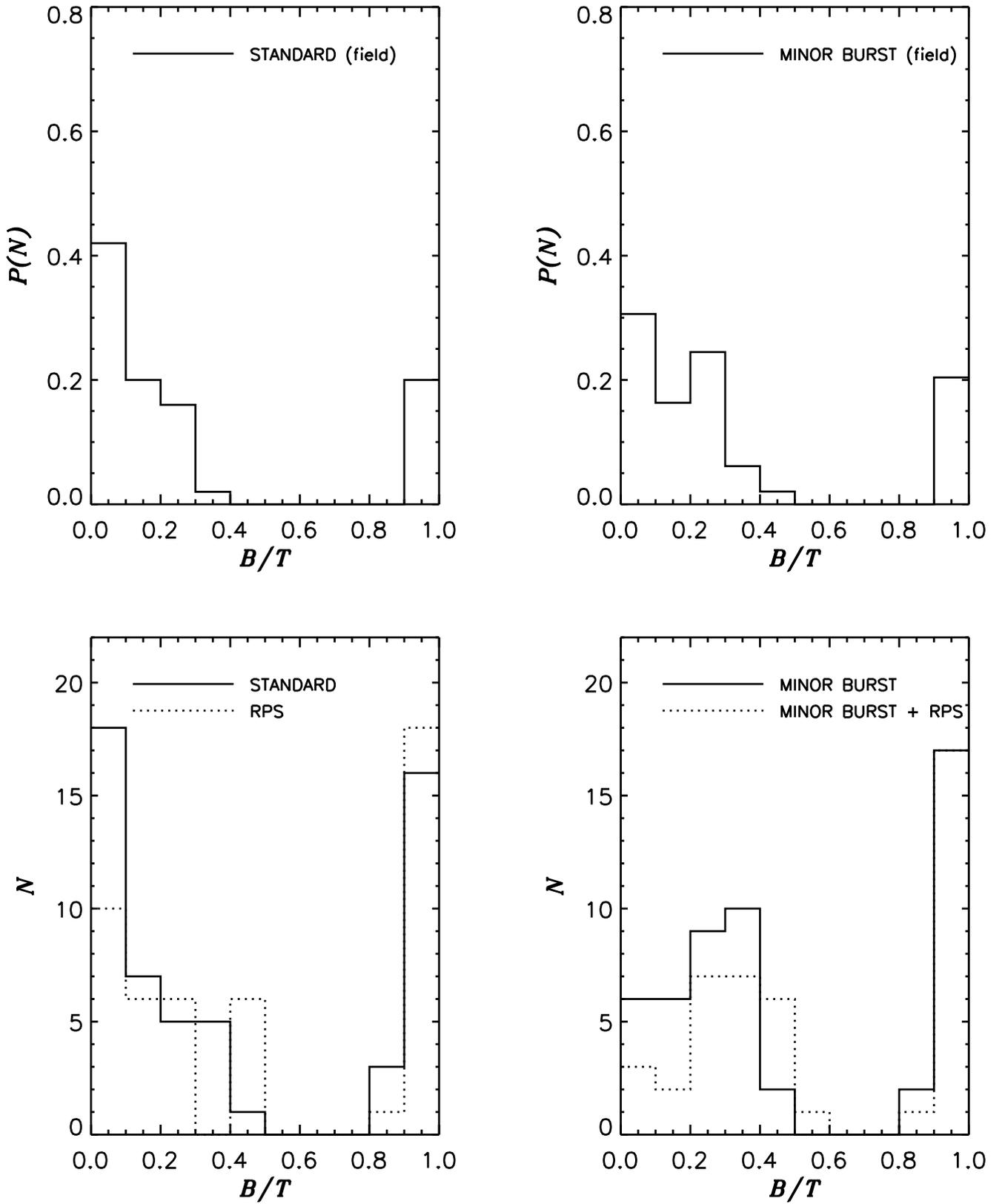}
\caption{
The distribution of $B$-band $B/T$ in the field (upper panels) and cluster (lower panels). 
we show the probabilistic distribution for the field galaxies and the number distribution for the cluster galaxies. 
In the lower panels the corresponding RPS modes are also shown by the dotted lines. 
\label{fig8}}
\end{figure}

To clarify the effects of the minor burst and RPS, we show $B/T$
distributions for our model galaxies with $M_R < -20.5$ at $z = 0.2$ in
Fig.~\ref{fig8}.
The probabilistic number distributions for the field
are represented as a function of $B/T$ in top panels. In the standard
model, there is a large blank between $B/T = 0.4$ and 0.9, and the
distribution has peaks at $B/T \simeq 0$ and 1.  In the minor burst model
the galaxies have similar $B/T$ distribution to that in the standard
model, while the peak at $B/T \simeq 0$ is lowered and the galaxies have
slightly extended $B/T$ distribution.

In the lower panels we show the $B/T$ distributions of the galaxies
within $R_{200}$ in the standard models (left) and the minor burst
models (right). The models with and without the RPS are indicated by the
dotted and solid lines, respectively. To see the
influence of the RPS, we do not normalize the distribution by the total
number of galaxies.

In the standard model, the $B/T$ distribution has peaks at $B/T \simeq
0$ and $1$ as well as in the field model.  By comparing the standard
model to the RPS model, it is found that the RPS changes the
distribution in two ways.  Firstly, it decreases the number of galaxies
brighter than the luminosity cut-off.  Since the RPS only affects the
luminosity of the disk component, the number of the disk-dominated
galaxies is more easily decreased than the galaxies with large $B/T$s.  
Secondly, the RPS increases the $B/T$s of non-pure disk (and
of course non-pure bulge) galaxies. In our case, this effect is the most
significant around $B/T = 0.4$.

In the minor burst model, lower $B/T$ distribution is significantly
changed from that in the standard model. The peak at $B/T \simeq 0$ in
the standard model moves to $B/T = 0.3-0.4$ in the minor burst
model. In the cluster the effect of the minor burst is stronger than in the
field because of the higher merger rate.  The RPS works in the same way as
in the standard model, but now the galaxies in the cluster center have
larger bulges than those in the standard model (Fig.~\ref{fig7}), and
then the RPS increases the intermediate $B/T$ fraction in the center.

As we have shown here, the $B/T$ distribution gives us much more useful
information than classifying galaxies into three types. Therefore the
direct comparison of the $B/T$ distribution predicted by a model with 
observational one in various environments will be a crucial test for the
morphology evolution model.

\subsubsection{Morphology evolution} \label{morph-z}

\begin{figure}
\epsscale{0.8}
\plotone{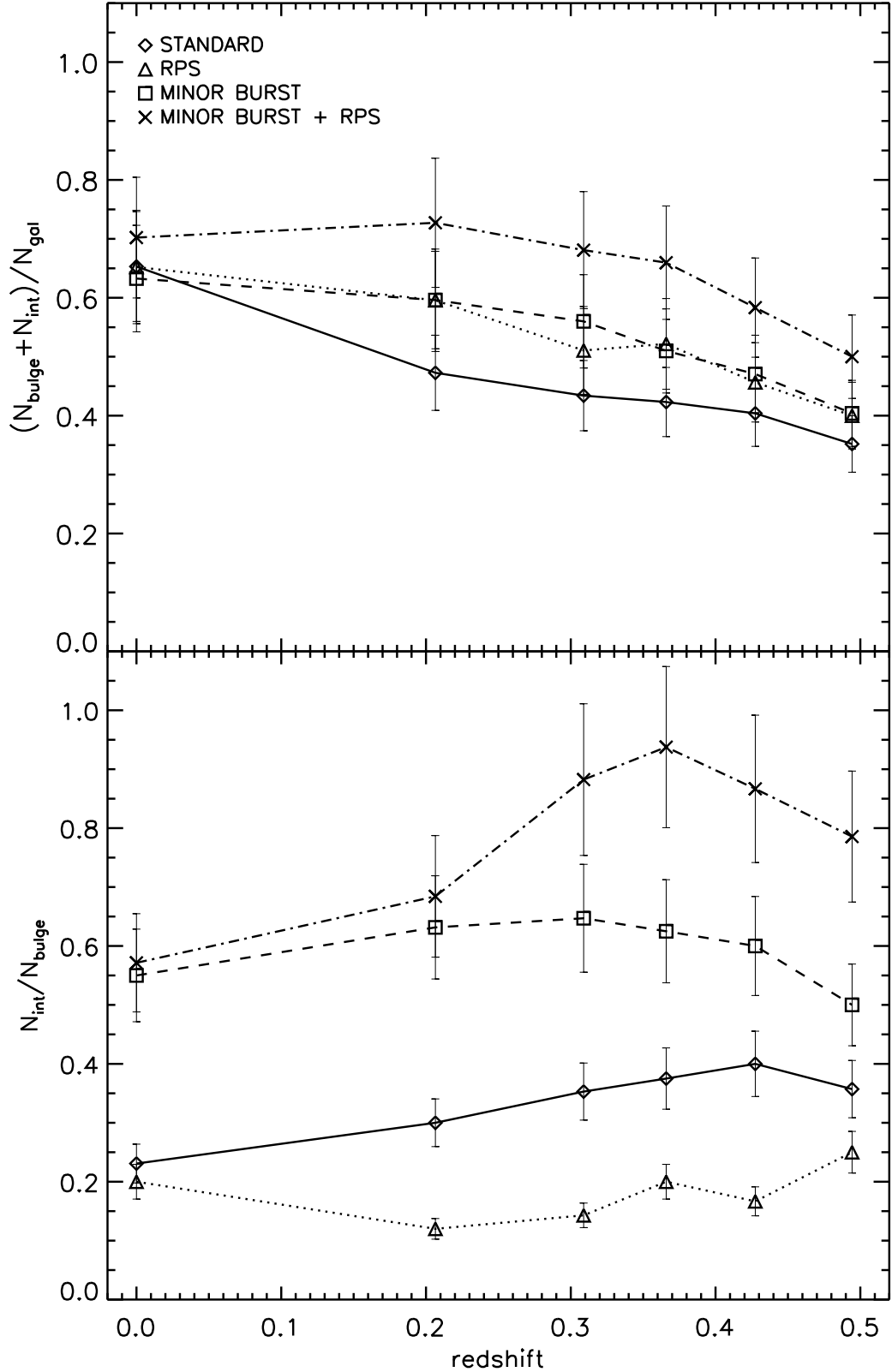}
\caption{
{\it Upper}: The non-disk-dominated fractions in the cluster virial radius as a function of redshift. The diamonds, triangles, squares, and crosses indicate the standard, RPS, minor burst, and minor burst + RPS models, respectively. The classification criteria are the same as Fig.~\ref{fig7}. 
{\it Lower}: Int/Bulge-ratio as a function of redshift. 
The error bars present the Poisson errors. 
\label{fig9}}
\end{figure}

\citet{dre97} noted that the S0 fraction of cluster galaxies appears to
decrease with redshift.  Although \citet{and98} argued that it is
difficult to define a reliable S0/E-ratio, \citet{vanD01} pointed out
that, in clusters, at least the early-type (E+S0) fraction decreases
with redshift. To probe the role of each environmental effect in the
morphology evolution we plot the non-disk-dominated galaxy fractions
and the number ratios of the intermediate galaxies to the
bulge-dominated galaxies (hereafter Int/Bulge-ratios) in the upper and
lower panels of Fig.~\ref{fig9}, respectively. We select the galaxies
brighter than $M_R = -20.5$ within $R_{\rm vir}$ at each redshift.

All models show the gradually decreasing non-disk-dominated fractions
with redshift. This qualitatively agrees with the observational trend of
decreasing early-type fraction in clusters with redshift, though we
cannot make quantitative comparison because the correlation between the
Hubble-type and $B/T$ (Eq. (\ref{morphology})) has a quite large scatter
and we do not know whether this relation is adequate for cluster
galaxies. There is no doubt that the hot gas stripping and the RPS is
responsible for the change of the non-disk-dominated fractions. These
processes not only increase the $B/T$ of the galaxies but also decrease
the number of galaxies brighter than the cut-off luminosity. The latter
is also important because the disk-dominated galaxies are more easily 
darkened by these stripping processes, and then the disk-dominated
fraction decreases toward lower redshift.

Our Int/Bulge-ratios show completely different trends from the observed
S0/E-ratio.  These ratios are flat or moderately increasing functions of
redshift, albeit the observed S0/E-ratio is a strongly decreasing
function.  This feature can be understood from the following reason.  
In our models the bulges are formed through mergers and the merger rate is
only high before cluster formation.  Thus, the bulges hardly grow up in
clusters.  Hence, the main process that evolves the galaxy
morphology in clusters is the fading of disks through gas stripping. As
we mentioned above, this process increases $B/T$ (i.e. it increases the
bulge-dominated and intermediate galaxies) and darkens the galaxies.  As
a result, although both the bulge-dominated and intermediate galaxies
are increased, the intermediate galaxies are more likely to be darkened
and fall below the cut-off luminosity than the bulge-dominated
galaxies. Then the Int/Bulge-ratios in the cluster become flat or
increasing functions of redshift under the merger-driven bulge formation
model. 

We have to consider the processes that form or evolve bulges in clusters
in order to produce an Int/Bulge-ratio similar to the observed
S0/E-ratio.  However, it is probably still possible that the
Int/Bulge-ratio shows different evolution from the S0/E-ratio because
the eyeball classification does not depend only on the $B/T$.  In fact,
the disks of luminous Sa-Sb galaxies are substantially thickened and the
spiral features are vanished by tidal encounters in clusters and they become
galaxies resembling S0's without the large change in their $B/T$s
\citep{moo99}.  We expect that observations of the Int/Bulge-ratio as a
function of redshift will give important information about the formation
process and/or epoch of the intermediate $B/T$ population.

\section{SUMMARY AND DISCUSSION}

In this paper, we have investigated the effects of hot gas stripping,
RPS of cold disk gas, major mergers, and minor mergers on the evolution of
bright cluster galaxies, for which the other mechanism, say, 
galaxy harassment, is probably not the predominant process.
We have used a combination of the cosmological $N$-body
simulations and the SA galaxy formation model.  This
method enables us to study above environmental effects in a fully
cosmological context.  We have determined the model parameters of the
reference model to reproduce observations of nearby field galaxies and
CMRs of local cluster ellipticals.

Our main results are summarized as follows.
\begin{itemize}
	\item The process that terminates star formation in early-type
	      galaxies is starburst, because the CMRs for early-type galaxies in the
	      cluster are almost independent of either the star formation 
	      time-scale or the presence of the RPS. 
	\item If we adopt appropriate star formation time-scale,
	      so as to reproduce the observed cold
	      gas mass fraction in the field, the dominant process that
	      determines colors of galaxies in the cluster core is the
	      stripping of hot gas reservoirs. The effect of the RPS is
	      only substantial when we adopt much longer star formation
	      time-scale.
	\item Minor burst does not affect galaxy properties except for
	      morphology.
	\item Without minor burst galaxies typically separate into
	      almost pure disk ($B/T \simeq 0$) and almost pure bulge
	      ($B/T \simeq 1$) galaxies. The gap in the $B/T$ distribution
	      causes the deficiency of the intermediate $B/T$ galaxies
	      and makes the intermediate galaxy fraction very sensitive
	      to the adopted $B/T$ range. The minor burst decreases the
	      galaxies with $B/T \simeq 0$ and increases those with $B/T
	      = 0.3-0.4$. This effect is larger in the cluster than
	      in the field reflecting a richer merger history, and then
	      the model with the minor burst can reproduce the observed
	      morphology-radius relations.
	\item RPS gives quite small changes in the morphological
	      fractions in the central part of the cluster under the
	      bulge formation only by major mergers. When the minor
	      burst is considered, the disk-dominated galaxies in the
	      cluster center tend to have larger bulges than in the
	      standard model, so that the RPS increases/decreases the
	      intermediate/disk-dominated galaxy fraction there.
	\item The fractions of the non-disk-dominated galaxies in the
	      cluster are decreasing functions of redshift. This feature
	      agrees with the observed evolution of early-type fraction
	      in clusters. On the other hand, the Int/Bulge-ratios are
	      flat or slightly increasing functions of redshift unlike
	      to the observed S0/E-ratio.
\end{itemize}

The stripping of the hot gas from galactic halos suppresses the star
formation in the cluster core so efficiently that the effect of RPS is
seen only at the outskirts of the cluster.  In order to find out this
effect observationally, it would be worthwhile comparing the colors or
SFRs of the galaxies in cluster outskirts with those of the galaxies
lying in regions with the same local density as the cluster outskirts
but not associated with clusters.  Note that we do not have enough
samples at the cluster outskirts and we do underestimate field
contamination because of the small volume and spherical shape of the
high resolution region in our simulation. Moreover we assume
instantaneous stripping of cold gas by the RP as a maximal effect, so that we
must overestimate the stripping effect. Thus, it would be more difficult
to see the difference by RPS in observations.

We have shown that the bulge formation solely by major mergers has the
difficulty to form galaxies of intermediate $B/T$ ($\sim 0.3-0.8$) both 
in the field and cluster. The minor burst can resolve this problem because 
it develops the bulge of the disk-dominated galaxies.  
This effect is more significant for the cluster galaxies than the field
galaxies reflecting richer merger histories. Then the minor burst can
reproduce the observed morphology-radius relation. Since the disk-dominated
galaxies in cluster tend to have larger bulges than those in the field
by the minor burst, it also gives a reasonable answer to the problem
that the fading of the disk-dominated galaxies like field spirals cannot
explain the entire cluster S0 population as there is too little
luminosity density in spiral bulges \citep[e.g.][]{dre80}. When the
minor burst is considered, the RPS can increase/decrease the
intermediate/disk-dominated galaxy fraction at the cluster center, while
the colors and SFRs of these galaxies are hardly changed.  
Our results show that the minor burst is very important process to explain
morphology evolution of galaxies, although our treatment is too
simple. In order to construct a feasible morphology evolution model
induced by mergers, the comparison of model predictions and observations in
low density environments (field and groups) where the RPS and the galaxy
harassment are not important and the merger is predominant process that
leads morphological transformation will be quite helpful.

In all our models the bulge formation is driven by mergers, and then the
Int/Bulge-ratios become flat or slightly increasing functions of
redshift contrastively to the observed S0/E-ratio in spite of we include
the fading of disks by the hot gas and/or cold gas stripping. This is
because the merger rate is only high before the cluster formation
\citep[][]{oka99,got01} and the bulges hardly evolve in the cluster.
This discrepancy between our models and the observation can be
explained by the difference in the classification methods, that is, the
Hubble type of a galaxy depends on its SFR as well as its $B/T$.  
Consequently star-forming intermediate $B/T$
galaxies could be classified as spirals and non-star-forming
disk-dominated galaxies have a chance to be classified as S0's.  
If the Int/Bulge-ratio is also observed as a decreasing function of
redshift, we will have to consider other bulge formation processes that 
increase bulge luminosity of the disk-dominated galaxies in high density
environment.

\acknowledgments

We are grateful to S. Mineshige, Y. Fujita, and T. Kodama for their 
fruitful comments. 
Numerical computation in this work was carried out on SGI Origin 
2000 at Division of Physics, Graduate School of Science, 
Hokkaido University and on SGI Origin 3000 at Yukawa Institute 
Computer Facility.


\clearpage





\clearpage

\begin{deluxetable}{ccrrrrrrrr}
\tabletypesize{\scriptsize}
\tablecaption{Parameters of Simulations \label{tbl-1}}
\tablewidth{0pt}
\tablehead{
\colhead{Simulation} & \colhead{Constraint}   & \colhead{$N_{\rm h}$}   &
\colhead{$N_{\rm l}$} & \colhead{$\epsilon_{\rm h}$}   & 
\colhead{$\epsilon_{\rm l}$} & 
\colhead{$m_{\rm h}$} &
\colhead{$m_{\rm l}$} &
\colhead{$R_{\rm sim}$}  &
\colhead{$M_{\rm largest}$} 
}
\startdata
Field &None &636008 &824438 &5e-3  &2e-2 &5.5e+8  &4.4e+9 &20 & 
1.87e+13  \\
Cluster &$3 \sigma$ peak &1121534 &95406 &5e-3 &5e-2 &5.5e+8 &3.5e+10 &22.5 &5.23e+14  \\
\enddata
\tablecomments{
Parameters $N$, $\epsilon$, $m$, $R_{\rm sim}$, and $M_{\rm largest}$ 
represent number of particles in each simulation, mass of a particle, radius 
of each simulation sphere, and mass of the largest virialized object 
in each simulation at $z = 0$, respectively.
Subscripts h and l indicate high resolution and low resolution particles, 
respectively. 
A length unit is $h^{-1}$ Mpc and a mass unit 
is $h^{-1} \ M_{\odot}$}. 
\end{deluxetable}


\begin{deluxetable}{cccccc}
\tabletypesize{\scriptsize}
\tablecaption{Parameters of Galaxy Formation Model \label{tbl-2}}
\tablewidth{0pt}
\tablehead{
\colhead{$V_{\rm hot}$ (km s$^{-1}$)}  &  \colhead{$\alpha_{\rm hot}$}  & 
\colhead{$\tau_*^0$ (Gyr)}  &  \colhead{$\alpha_*$}   & 
\colhead{$f_{\rm major}$}  &  \colhead{$y$ ($Z_{\odot}$)}  
}
\startdata
200 & 2.0 & 2.0 & -1.3 & 0.2 & 3.0 \\
\enddata
\end{deluxetable}


\clearpage 

\clearpage 

\clearpage 

\clearpage 

\clearpage 

\clearpage 

\clearpage 

\clearpage 

\end{document}